# The β to γ (insulator-metal) transition in BiFeO$_3$

Donna C. Arnold[1], Kevin S. Knight[2], Gustau Catalan[3], Simon A.T. Redfern[3], James F. Scott[3,4], Philip Lightfoot[1] and Finlay D. Morrison[1*]

[1] *School of Chemistry, University of St Andrews, North Haugh, St Andrews, Fife, KY16 9ST, UK*

[2] *ISIS Facility, Rutherford Appleton Laboratory, Chilton, Didcot, OX11 0QX, UK*

[3] *Department of Earth Sciences, University of Cambridge, Downing Street, Cambridge, CB2 3EQ, UK.*

[4] *Department of Physics, University of Cambridge, J. J. Thomson Ave., Cambridge, CB3 0HE, UK*

[*] *To whom correspondence should be addressed: e-mail: fm40@st-andrews.ac.uk, Tel: +44-1334463855, Fax: +44-1334463805.*

**Abstract**

High temperature powder neutron diffraction experiments have been conducted around the reported β-γ, insulator-metal phase transition (~ 930 °C) in BiFeO$_3$. The results demonstrate that while a small volume contraction is observed at the transition temperature, consistent with an insulator-metal transition, both the β- and γ-phase of BiFeO$_3$ exhibit orthorhombic symmetry i.e. no further increase of symmetry occurs under the present experimental conditions, contrary to previous suggestions. Furthermore we observe the γ-orthorhombic phase to persist up to a temperature of approximately 950 °C before complete decomposition into Bi$_2$Fe$_4$O$_9$ (and liquid Bi$_2$O$_3$), which subsequently begins to decompose at approximately 960 °C.

# 1. Introduction

The drive to find novel materials for applications such as data storage, spintronics and microelectronic devices has lead to a resurgence in research into multiferroic materials.[1,2] By far the most widely studied multiferroic material is $BiFeO_3$, due primarily to its high temperature magnetic and electric transition temperatures: ferroelectric $T_C \sim 810 - 830$ °C[3,4] and antiferromagnetic $T_N \sim 350 - 370$ °C.[5]

Despite extensive studies on bulk $BiFeO_3$ there are still many contradictions in the literature.[6] In particular the high temperature phases have been well studied, though are still not fully understood. In the early work as many as eight anomalies in physical properties as a function of temperature, based on electrical and magnetic measurements, have been reported.[7,8] However, it is now more generally accepted that there are two main anomalies which coincide with the transitions at $T_N$ and $T_C$. A third transition has been reported by some authors at a temperature of approximately 190 °C.[8] However, the nature of this transition has never been fully understood, with some authors reporting evidence for this transition in structural studies whilst others make no reference to it. Variable temperature X-ray powder diffraction (XRPD) studies[9], suggested two anomalies in lattice parameters at $T_N$ and $T_C$ and this was later supported by a powder neutron diffraction (PND) study.[10,11] These authors reported the evolution of the rhombohedral (R3c) α-phase up to a temperature of 700 °C before significant decomposition to $Bi_2Fe_4O_9$ occurred; a problem which occurs in many studies of $BiFeO_3$ at elevated temperatures.

More recently an updated phase diagram for $BiFeO_3$[12] has been reported based on thermal analysis, spectroscopic, diffraction and other methods. This work suggests evidence of three distinct solid phases above room temperature and below the melting point (~960 °C): the rhombohedral α-phase, below $T_C$, an intermediate β-phase, in the region 830 – 925 °C, and a cubic γ-phase in the region 925 – 933 °C before decomposition and subsequent melting. Many symmetries for the β-BiFeO$_3$ phase have been suggested by x-ray diffraction and first principle techniques including tetragonal (space group I4/mcm),[13] monoclinic (P2$_1$/m),[14] orthorhombic[12,15] and rhombohedral (R-3c).[16] Recently we demonstrated, by high resolution PND experiments, that the β-phase crystallises in the orthorhombic Pbnm space group, the same as non-polar GdFeO$_3$.[15]

In contrast, little work has been reported for the β-γ phase transition, primarily due to the onset of decomposition before this temperature is reached.[14,15] Palai *et al.* suggested the γ-phase adopts cubic symmetry, with this transition reported to coincide with a insulator-metal transition.[12] This was supported by band structure models, high-temperature dc resistivity and light absorption and reflectivity measurements which show that the band gap decreases with increasing temperature before dropping abruptly to zero at ~930 °C.[12] More recently the presence of a metal-insulator transition has been reported using ab-initio calculations[17] and in high pressure[18,19] and spectroscopic ellipsometry studies.[20,21]

Here we present high temperature powder neutron diffraction data in the temperature range 900 °C to 960 °C. Through careful design of the experimental set-up and the collection of short data runs (~5 mins) we have been able to minimise the onset and subsequent rate of decomposition of BiFeO$_3$ into the parasitic Bi$_2$Fe$_4$O$_9$ phase. We report a subtle drop in the cell volume at approximately 930 °C consistent with the previously reported insulator-metal transition. However, in contrast with previous reports we demonstrate that BiFeO$_3$ is in fact orthorhombic over the whole temperature range investigated. This has important consequences for our understanding of the origin of the metal-insulator transition.

## 2. Experimental

Single phase $BiFeO_3$ was prepared using the method previously described by us.[15] Stoichiometric ratios of $Fe_2O_3$ (Aldrich, ≥ 99 %) and $Bi_2O_3$ (Aldrich, 99.9 %) were reacted with a 6 mol % excess of $Bi_2O_3$ at 800 °C for 5 hours. After calcination the material was ground into a fine powder and leached with 2.5 M nitric acid under continuous stirring for 2 hours, washed thoroughly with distilled water and dried at 400 °C for 1 hour. XRPD confirmed single phase $BiFeO_3$.

PND data were collected on the HRPD instrument at ISIS, over a temperature range of 900 °C to 960 °C. Short data collections (~5 mins) were undertaken when the furnace reached temperature, with a second run (~ 5 mins) collected immediately after the first. A third 5 minute run was also collected at 900 °C in order to allow the material to equilibrate at these elevated temperatures. Due to the high partial pressures in the system as a result of the high temperature decomposition of $BiFeO_3$ into $Bi_2Fe_4O_9$ and liquid $Bi_2O_3$ the experiment was conducted under flowing $N_2$. This was achieved by taking $BiFeO_3$ powder compacts which were suspended on a quartz frit within a quartz tube. The tube was then mounted within the furnace with the thermocouples attached to the top of sample material and $N_2$ flowed over the closed system.

For XRPD data collection polycrystalline ceramic $BiFeO_3$ was ground under acetone and mounted on an alumina sample holder within the MRI high-temperature chamber of a Bruker D8 Advance diffractometer. Data were collected in $\theta$–$\theta$ geometry using CuK$\alpha$ radiation, Göbel mirror parallel beam primary optics and Soller receiving slits with a Vantec position sensitive detector. Diffraction patterns were obtained in situ at high temperature by rapid scans over the region of interest (20˚ to 60˚ 2$\theta$) such that the sample was held at each temperature for no more than five minutes, with temperature steps varying between 15 and 20 K.

3. Results and Discussion

Crystallographic analysis of PND data collected at 900 °C confirmed the formation of β-$BiFeO_3$. Interestingly, as a result of the fast heating rates and short scan times employed in order to minimise decomposition to $Bi_2Fe_4O_9$ and liquid $Bi_2O_3$, a small amount of the α-phase (i.e. R3c phase) was seen to persist for approximately 10 minutes at 900 °C before full transformation into the Pbnm β-phase. Data collected for the 3$^{rd}$ five minute run showed no evidence of any un-indexed peaks further confirming our previous assessment of the orthorhombic character of the β-phase.[15] Rietveld refinement of these data was performed using the GSAS suite of programs[22] with the refinements including individual isotropic atomic displacement parameters for all atoms as well as the usual profile coefficients. The refinement profile is given in figure 1 with further details given in table 1. With increasing temperature $BiFeO_3$ can be seen to decompose to $Bi_2Fe_4O_9$ (and liquid $Bi_2O_3$), with approximately 5 % of this phase observed at 910 °C, increasing rapidly to approximately 80 % at 950 °C despite the short scan times and high heating rates employed. The degree of decomposition of $BiFeO_3$ increased steadily with temperature as shown in supplementary information, figure S1. A reliable full refinement was also performed on the 2nd 945 °C data set as given in figure 1(b) and table 1. However at intermediary (and higher) temperatures, whilst all data were modelled using the Rietveld method, it was only possible to extract quantitatively the lattice parameters as a result of increasing amounts of 2$^{nd}$ phase and in the intermediary data the presence of shoulders on the main BiFeO$_3$ peaks (discussed below). A summary of the extracted parameters and relative phase percentages over the temperature range 910 to 950 °C can be found in supplementary data, table S1.

**3.1 Volume contraction**

The thermal evolution of the cell volume of the BiFeO$_3$ phase is shown in figure 2. Between 900 °C and initial data collected at 930 °C the cell volume continues to increase as expected from thermal expansion and in agreement with data collected previously by us (figure 2b).[15] During data collection at 930 °C a small decrease in the cell volume (ΔV/V = 0.15 %) is observed between runs 1 (0-5 min) and 2 (5-10 min). On continued heating the cell volume stabilises at values less than those observed between 920 and the initial run at 930 °C. This discontinuous volume change is consistent with a first order phase transition and the difference observed between runs 1 and 2 at 930 °C is due to a region of phase co-existence associated with a first order transition, and shows the kinetics of the transformation process. The drop in cell parameters and cell volume is likely to result from shortening of the Fe-O bonds as the phase becomes metallic.

These observations are consistent with the optical observations and band gap calculations of Palai *et al*.[12] who were the first to suggest an insulator-metal transition. This subtle change in volume has also been observed at the metal-insulator transition in other transition metal perovskite oxides such as NdNiO$_3$.[23] It would also be expected that this transition would be accompanied by a straightening of the Fe-O-Fe bond angle. However, it was not possible perform full refinements to obtain accurate atomic positions and hence bond angles over the whole temperature range due to the deterioration in data quality due to the increased temperature and also the decreasing signal-to-noise as the BiFeO$_3$ phase decomposes.

**3.2 Symmetry of the γ-phase**

Previous reports have suggested that the metallic phase (γ-phase) exhibits cubic, Pm-3m, symmetry.[12] This orthorhombic-cubic transition has also been reported in high pressure experiments on BiFeO$_3$[18] and also in Mn-doped samples (BiFe$_{0.7}$Mn$_{0.3}$O$_3$) at ambient pressure, which exhibits an analogous R3c to Pbnm to Pm-3m series of phase transitions.[24] However, the latter remains semiconducting throughout with no evidence of a metal-insulator transition. In fact in this material the transition from Pbnm to Pm-3m is marked by a sudden increase in cell volume as opposed to the volume shrinkage expected in an insulator-metal transition and observed in pure BiFeO$_3$. Close inspection of our data provides no evidence for the existence of a cubic phase, with peak splitting and peaks which violate cubic symmetry remaining clearly evident right up to full decomposition and melting (figure 3). Considering the work of Howard and Stokes[25] the most likely tilt arrangements for perovskites transforming from Pbnm (Glazer notation a$^-$a$^-$b$^+$) towards Pm-3m (a$^0$a$^0$a$^0$) symmetry would suggest a possible transition to either tetragonal, P4/mbm (a$^0$a$^0$c$^+$) or orthorhombic, Imma (a$^0$b$^-$b$^-$) symmetry. As with the cubic symmetry the presence of peaks and peak splitting which violate tetragonal symmetry rule out the possibility of a transition to tetragonal, P4/mbm, symmetry. This is particularly evident in the d-spacing region 2.20 – 2.40 Å, where two peaks (assigned as (202) and (022) in orthorhombic, Pbnm, symmetry) are clearly apparent over the whole temperature range investigated; these would be required to be a single peak to satisfy either tetragonal or cubic symmetries, figure 3(d). The possibility of an orthorhombic, Pbnm, to orthorhombic, Imma, phase transition cannot be ruled out on this basis since both phases would result in the two peaks observed.

The data were more closely scrutinised in an attempt to confirm or discount the presence of any Imma phase by looking for the presence of peaks which are present in Pbnm but are systematically absent in Imma. Unfortunately the most intense of these peaks, including the (212/122) reflections, are masked by the increasing amount of $Bi_2Fe_4O_9$. The data do, however, show reflections at d-spacing *ca.* 2.52 Å corresponding to Pbnm (210/120) and which are systematically absent in Imma (see figure S2 in supplementary information). However, the intensity of these peaks decreases such that they are barely evident above 930 °C and so an Imma phase cannot be conclusively excluded. These results however do clearly and unambiguously indicate that both sides of the metal-insulator phase transition exhibit orthorhombic symmetry, such that either both the insulating and the metallic phase exhibit Pbnm symmetry, or possibly a Pbnm to Imma phase transition occurs. These observations are remarkably similar in nature to the case of orthorhombic $RNiO_3$ (R = Nd or Pr)[23], which exhibit metal-insulator (semiconductor) transitions which are not accompanied by a change of symmetry i.e. both phases exhibit Pbnm symmetry.

Further inspection of data between 900 °C and 940 °C reveals the presence of weak shoulders on the main peaks of the Pbnm phase (figure 4), suggesting a possible region of phase co-existence either between the two orthorhombic phases or between the insulating and metallic phases, indicative of a first order phase transition. Again, this is remarkably similar to the $RNiO_3$ phase where the metal-insulator transition is first order in nature resulting in regions of phase co-existence.[26] Above 945 °C these shoulders have disappeared completely and the data gives a good fit to the Pbnm model, suggesting that full transformation of the phase has occurred, further confirming that both the β- and γ-phases exhibit orthorhombic character in contrast with previous reports.[12,18] Attempts at multiphase refinements to model possible phase co-existence of an orthorhombic β-phase with either a cubic or tetragonal γ-phase were unsuccessful with both models failing to fit all shoulders present. The possibility of phase co-existence between two orthorhombic (Pbnm) phases with different lattice parameters was also modelled. Figure 4 clearly shows excellent fitting of the shoulders further supporting the suggestion that these shoulders arise as a result of a first order transition between two orthorhombic (Pbnm) phases. The presence of such shoulders do not significantly affect the lattice parameters obtained for the orthorhombic Pbnm phase and again, data at 945 °C do not show any evidence of shoulders and give smaller lattice parameters and cell volume reduced than at temperatures in the range 910-930 °C. On a cautionary note, whilst the evidence supports a first order transition between two phases exhibiting orthorhombic symmetry, it should be stated the possible co-existence of an alternative orthorhombic phase (particularly Imma) can not be completely ruled out.

The γ-phase then persists up to temperatures of approximately 955 °C before complete decomposition occurs for the heating rates and dwell times used for data collection in these experiments. At this temperature (and above) the $Bi_2Fe_4O_9$ phase also begins to decompose into $Fe_2O_3$ and liquid $Bi_2O_3$ consistent with both the phase diagram and previous observations.[12,14]

### 3.3. Comparison with previous experiments

In an attempt to reconcile these observations with the previously reported cubic phase[12,18] we modelled the two symmetries (i.e. orthorhombic Pbnm and cubic Pm-3m) as x-ray diffraction patterns with peak profiles estimated from our previous XRPD data using the CrystalDiffract™ software[27]. Comparison of the predicted x-ray diffraction patterns are shown in figure 5 and clearly indicate the similarities between the two symmetries. In particular the FWHM of the reflection corresponding to orthorhombic (200)/(112)/(020) and cubic (110) peak is insufficiently narrow to allow resolution of any splitting and hence differentiation between orthorhombic and cubic symmetries. In fact only subtle differences can be observed in the form of weak superlattice reflections in the orthorhombic symmetry which are absent in the cubic symmetry. PND, however, shows clear differences as shown in supplementary information, figure S3. We have repeated our refinements of the XPRD data in light of the neutron results discussed above and these show adequate fits using the Pbnm model (figure 6). Furthermore extraction of the lattice parameters over the temperature range studied with XRPD showed excellent correlation with those extracted from the neutron data (see supplementary data, figure S4). This would suggest that the 'apparent' loss of the orthorhombic superlattice reflections with increasing temperature which we previously assigned as an orthorhombic-cubic phase transition occur as a result of the loss of instrument sensitivity.

These results highlight the limitations of laboratory-based x-ray diffraction studies for accurate phase identification in $BiFeO_3$ and emphasise the need for high resolution neutron experiments. Figure 7 shows the evolution of lattice parameters as a function of temperature as obtained from PND data.

**3.4. Does BiFeO₃ have a cubic phase?**

In addition to the conclusions that can be made directly from neutron scattering experiments, the domain wall structures in BiFeO$_3$ can be used to infer further information regarding structure and symmetry. J.-C. Toledano first demonstrated in 1974[28] that it is necessary and sufficient for ferroelastic phase transitions that the crystal undergo a change in crystal class (trigonal-hexagonal is regarded as a single super-class in this argument). Thus, rhombohedral-rhombohedral transitions, as in LiNbO$_3$, cannot be ferroelastic, nor can orthorhombic-orthorhombic transitions, as in KTiOPO$_4$. Ferroelectric but non-ferroelastic transitions can exhibit only 180° domains in the lower-symmetry phase. In BiFeO$_3$, however, the β–phase exhibits marked orthorhombic domains with walls that are NOT 180°: there are strong (110)pc walls ["pc" designates pseudo-cubic indexing], and weaker (100)pc walls intersecting them.[12] These non-180° walls therefore imply a higher-temperature phase (above the β-phase but not necessarily the γ-phase) that is not orthorhombic. This presents a paradox, since the present study shows clearly that the β-γ metal-insulator transition is orthorhombic-orthorhombic.

The resolution of this paradox is that there may be an additional phase higher in temperature than the γ-phase. There is no evidence that this phase is thermodynamically stable. However, even thermodynamically inaccessible states can produce domains and domain walls: a good example is the family BaMF$_4$ (M = Co, Ni, Mg, Zn, Mn, Fe), all of which melt before their extrapolated mm2 to mmm transition is reached; yet all of them form from the melt in multi-domain structures.[29]

More evidence for an 'inaccessible' cubic phase can be drawn from extrapolation of the pseudo-cubic lattice parameters of the orthorhombic β-phase. Such an extrapolation (figure 8) suggests that *a* and *b* parameters would converge at *ca.* 1020 °C, perhaps resulting in a tetragonal P4/mbm phase. Further extrapolation demonstrates that the lattice parameters converge to a single value above 1070 °C (figure 8); this would seem to support the existence of a high temperature transition to a non-orthorhombic (cubic, Pm-3m) symmetry. Such a sequence would follow the more widely observed sequence of transitions for perovskite oxides. While the extrapolation of data from the β-phase parameters should be viewed with caution, such a hypothesis is further supported by the Hamiltonians derived by Kornev *et al.* which predicted a transformation to cubic symmetry at a temperature of 1167 °C, well above the decomposition temperature.[13] In reality the onset of metallic character in the γ-phase results in a shortening of the Fe-O bonds and a drop in unit cell volume, which disrupts the "natural" structural progression towards cubic symmetry.

### 3.5. Metal-insulator transition

These observed results also have important implications for the nature of the metal-insulator (M-I) transition. Recently, different models for the metal-insulator transition, namely band-type[12] and Mott-type,[19] have been reported for $BiFeO_3$. Since a band-type insulator is required to have an even number of electrons in the unit cell (which fill the valence band completely such that there is a gap between them and the excited states in the conduction band) there has to be a structural transition whereby the number of formula units (and therefore the number of electrons) per unit cell changes in order for a valence band insulator to become a metal. On the other hand, in a Mott-type insulator the gap is due to electrostatic repulsion between the conduction electrons[30,31] and therefore and do not in theory require a change in either crystal structure or magnetic symmetry, although in practice the coupling between charge, spin and lattice means that other transitions tend to happen simultaneously.[32] The role of the electron-electron repulsion (the Mott-Hubbard parameter, U) in opening up the band-gap was discussed by Robertson and co-workers [12,33], and the challenge is to identify what is the mechanism that reduces this electron-electron correlation enough to drive the transition to the metallic state.

Previously it was believed that the metal-insulator transition in $BiFeO_3$ was driven by the structural transformation from orthorhombic to cubic symmetry which leads to an enhancement in orbital overlap between the iron and oxygen ions.[12,18] The data presented here clearly indicates that no structural transition occurs with both the β- and the γ-phases adopting orthorhombic symmetry. This would seem to rule out the possibility of a band-type model to describe the metal-insulator transition. It should also be noted that the possibility of a Pbnm to Imma phase transition could not conclusively be ruled out; however, both space group models exhibit identical unit cell parameters and therefore the same number of electrons which again rules against a band-type transition.

Gavriliuk *et al*. suggested a Mott-type transition, driven by a change in the spin state of the $Fe^{3+}$ ion from high spin to low spin, in a pressure induced phase transition,[19] although it unlikely that such spin transition model is also valid for the temperature-driven transition. Pressure-induced high-spin to low-spin transitions with associated decreases in resistivity have also been reported in other perovskite orthoferrites,[34,35] although the actual onset of the metallic state in these generally takes place at much higher pressures than the spin transition, indicating that the two are not necessarily correlated.

The absence of space group/symmetry change at a metal-insulator transition is not unheard of in perovskites. For example, it has previously been reported for the temperature-driven M-I transition in both $NdNiO_3$ and $PrNiO_3$, which exhibit orthorhombic Pbnm symmetry in both metallic and insulating phases[23,26] (and, interestingly, they have also been postulated to be multiferroic[26,36]). The M-I change in these compounds was postulated by Torrance *et al.*[37] to be caused by a critical straightening of the Ni-O-Ni bond angle, leading to increased orbital overlap between the oxygen *p* and nickel *d* orbitals and thus to a broadening of the conduction band across the Fermi level. On the other hand, very recent results suggest that the transition in $NdNiO_3$ and $PrNiO_3$ may in fact be neither isostructural nor due to a closing of the charge-transfer gap but to a charge disproportionation[38], so that the mechanism originally proposed by Torrance *et al.*[37] may in the end not apply to the nickelates themselves, although it might still apply to $BiFeO_3$[18, 20].

Another example is the pressure-induced spin-crossover transition in the rare-earth orthoferrites. They, like the β-phase of $BiFeO_3$, have Pbnm symmetry and, at high pressures (typically ~50Gpa) undergo a high-spin to low-spin transition in which only a small volume shrinkage and no symmetry change take place.[39] Moreover, at even higher pressures these rare-earth orthoferrites also become metallic[40] although it is presently unclear whether the metallisation is a direct consequence of the spin crossover[19,40] or due to a closing of the charge-transfer gap.[20]

**4. Conclusions**

In summary, we demonstrate that, in contrast with previous reports, both the β- and γ-phases exhibit orthorhombic symmetry either with no change in space group (Pbnm to Pbnm) or with a slight change in the octahedral tilt arrangement resulting in a transformation from Pbnm ($a^-a^-b^+$) to Imma ($a^0b^-b^-$). We also observe a subtle decrease in the unit cell dimensions during the transition which is consistent with the accompanying insulator-metal transition in $BiFeO_3$ (metallic bonds tend to be shorter due to electron delocalization).

The absence of space group/symmetry change at a metal-insulator transition has previously been reported in orthorhombic (Pbnm) nickelates.[23,26] These nickelates have been postulated to exhibit multiferroic character and therefore be fairly analogous to $BiFeO_3$, so that they may provide basis for a model to describe the transition in $BiFeO_3$. On the other hand, alternative models such as a high-spin low-spin crossover are also known to lead to isostructural transitions with associated volume shrinkage in orthoferrites, albeit so far only in pressure-dependent studies rather than temperature-dependent ones. A study of the spin behaviour of $BiFeO_3$ in the temperature-driven γ-phase will be required in order to ellucidate the ultimate cause of the M-I transition. Our results, meanwhile, prove unambiguously that the transition is not due to a structural change nor to a change in the number of electrons in the unit cell.

**Acknowledgements**

We thank the EPSRC, Royal Society and STFC for funding.

**List of Tables**

**Table 1:** Structural parameters for BiFeO$_3$ at 900 °C, Space group Pbnm, a = 5.6298(1) Å, b = 5.6536(1) Å, c = 7.9861(2) Å and cell volume = 254.19(1) Å$^3$, $\chi^2$ = 1.019, wRp = 7.97 %, Rp = 7.63 % and 945 °C, Space group Pbnm, a = 5.6311(3) Å, b = 5.6527(6) Å, c = 7.9878(3) Å and cell volume = 254.26(2) Å$^3$, $\chi^2$ = 0.9888, wRp = 8.20 %, Rp = 7.93 %.

| Atom | Site | 900 °C | | | | 945 °C | | | |
|---|---|---|---|---|---|---|---|---|---|
| | | x | y | z | U(iso) x 100 Å$^2$ | x | y | z | U(iso) x 100 Å$^2$ |
| Bi | 4c | 0.9993(9) | 0.017(1) | 0.25 | 7.08(9) | 0.997(5) | 0.021(6) | 0.25 | 9.3(6) |
| Fe | 4a | 0.5 | 0 | 0 | 1.32(3) | 0.5 | 0 | 0 | 3.2(3) |
| O | 4c | 0.0661(8) | 0.481(1) | 0.25 | 4.00(4) | 0.072(4) | 0.458(7) | 0.25 | 7.3(1) |
| O | 8d | 0.708(6) | 0.2925(6) | 0.0341(5) | 4.80(11) | 0.703(4) | 0.282(5) | 0.036(2) | 7.8(8) |

**List of Figures**

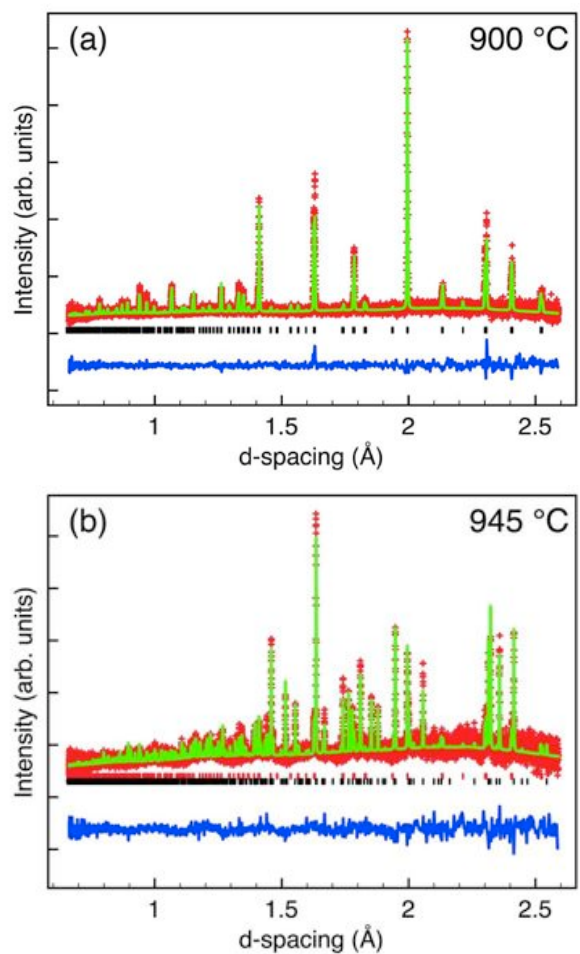

**Figure 1:** Refinement profiles at (a) 900 °C and (b) 945 °C showing the observed (+), fitted (solid line) and difference curves (bottom) for $BiFeO_3$ (both refined in the Pbnm space group) as represented by the bottom tick marks (top tick marks represent $Bi_2Fe_4O_9$). (Colour online).

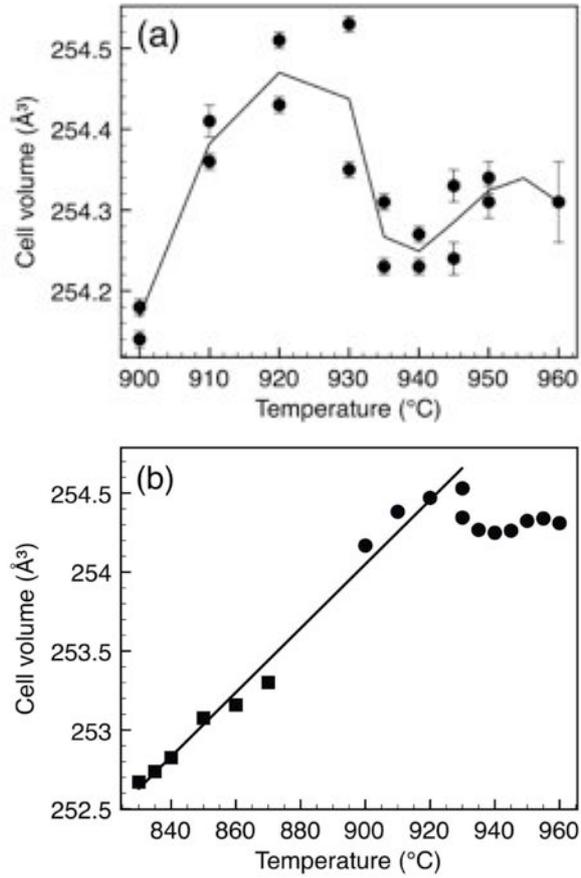

**Figure 2:** (a) Thermal evolution of the unit cell volume as a function of temperature in the β-BiFeO$_3$ phase. Data from runs 1 (0-5 min) and 2 (5-10 min) are shown at each temperature. The line indicates the average at each temperature and is provided as a guide to the eye. (b) Extrapolation to include data (■) collected in our previous experiments.[15]

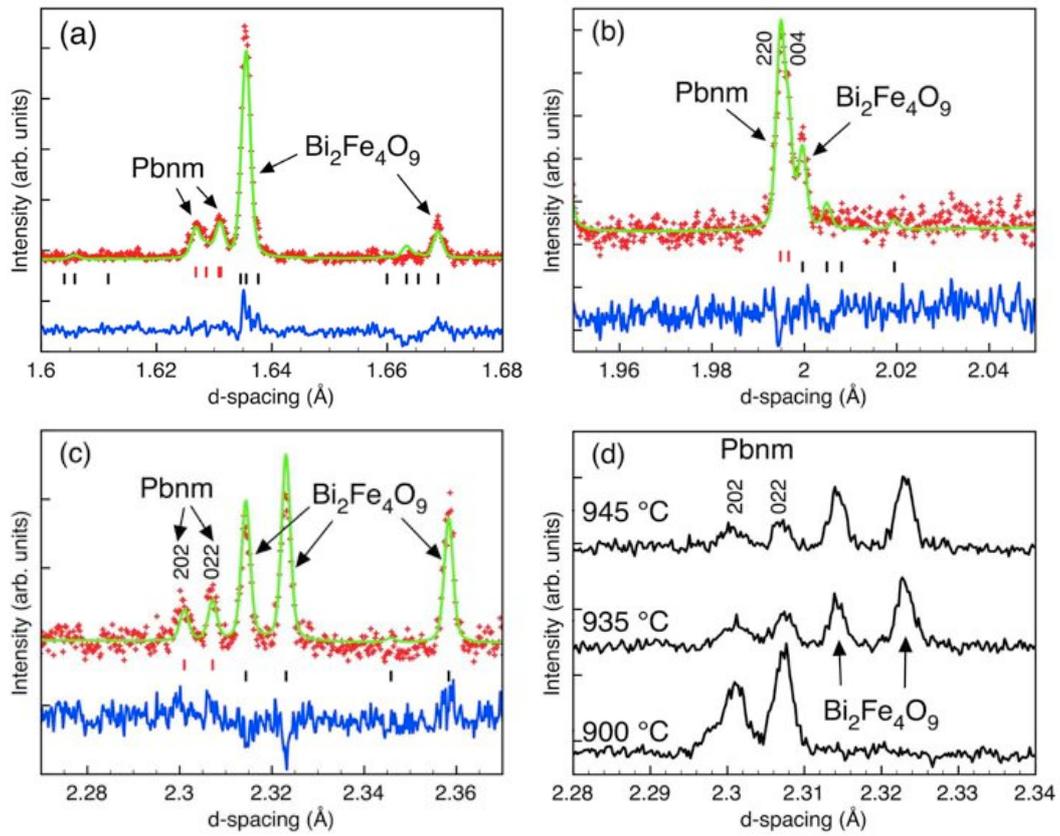

**Figure 3:** Expanded portions of the refinement profiles for data at 945 °C highlighting three different regions (a) to (c) clearly showing peak splitting which violate cubic and tetragonal symmetries and (d) showing that two distinct BiFeO$_3$ (202) and (022) peaks which violate cubic and tetragonal symmetries remain through the insulator-metal transition. (Parts (a)-(c) colour online).

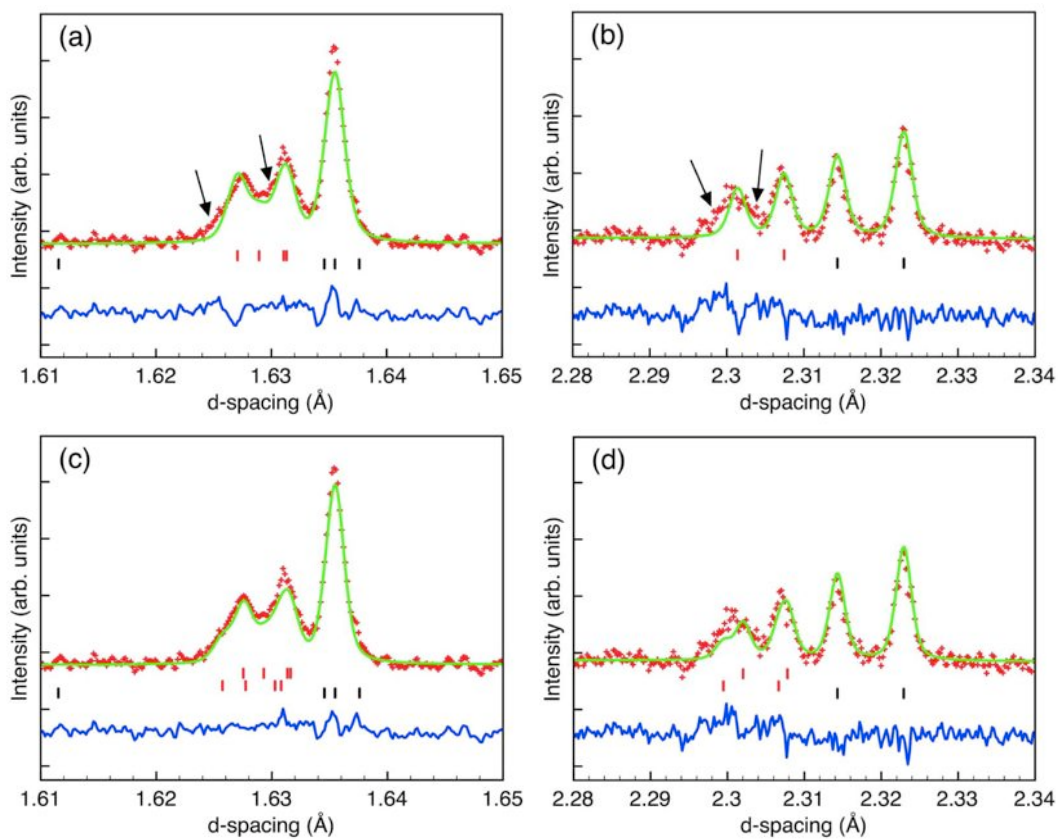

**Figure 4:** Expanded portions of the refinement profiles at 930 °C (a) and (b) fitted to a single Pbnm phase, clearly showing peak shoulders (marked by arrows) and the improved fit (c) and (d) achieved modelling two orthorhombic (Pbnm) phases. (Colour online).

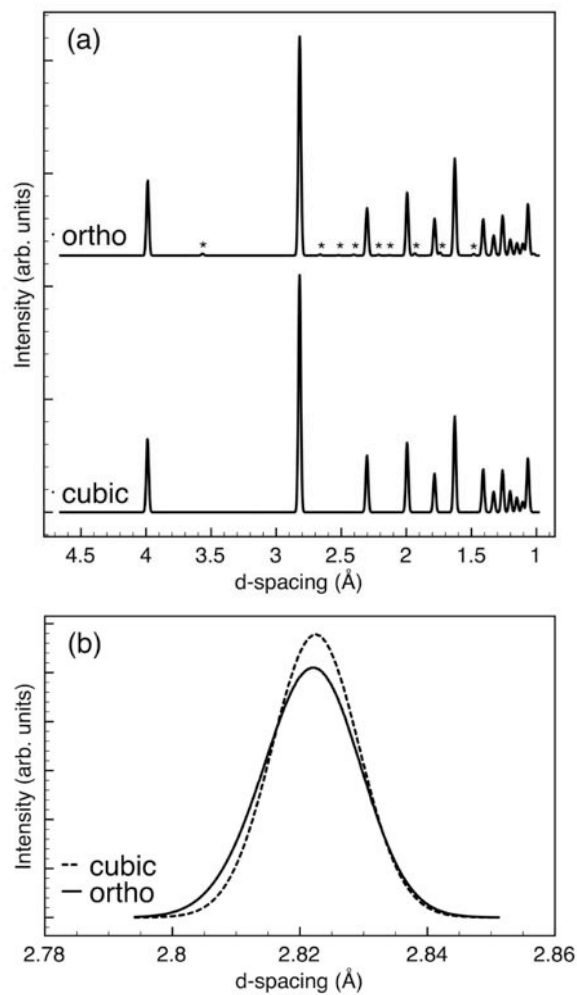

**Figure 5:** Simulation of the expected XRPD spectra (a) showing the differences between cubic (Pm-3m) and orthorhombic (Pbnm) symmetries emphasising the difficulties in resolving the weak additional reflections (marked *) associated with the orthorhombic symmetry and (b) expanded region indicating the similarities in the profiles of the cubic (110) and corresponding orthorhombic peaks with peak broadening as determined from data obtained at 950 °C (figure 6).

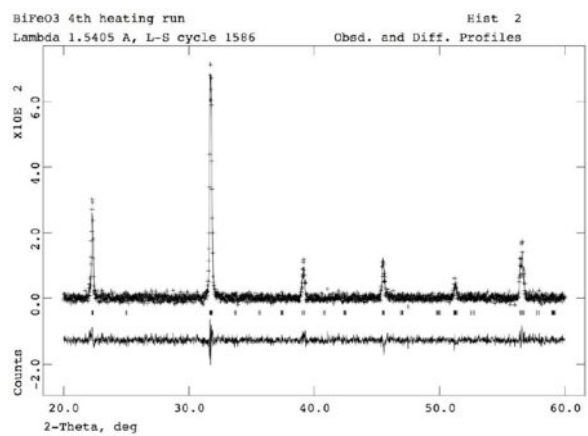

**Figure 6:** Refinement profiles of the XRPD data collected at 930 °C showing the observed (+), fitted (solid line) and difference curves (bottom) for $BiFeO_3$ (space group, Pbnm) as represented by the tick marks.

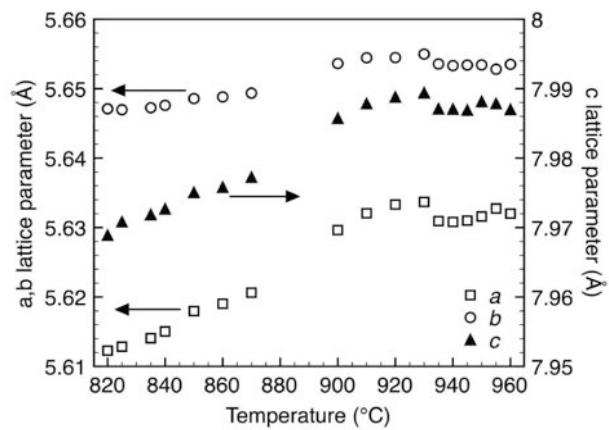

**Figure 7:** Lattice parameters as a function of temperature as extracted from powder neutron diffraction data. (Correlation between x-ray and neutron diffraction data is presented in supplementary information).

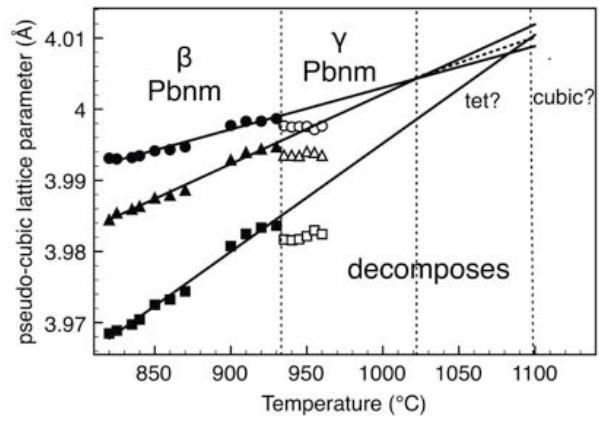

**Figure 8:** Extrapolation of pseudo-cubic cell parameters towards a unified value demonstrating the possible existence of an inaccessible cubic (or tetragonal) phase at temperatures beyond that of decomposition.

# The β to γ (insulator-metal) transition in BiFeO$_3$

Donna C. Arnold[1], Kevin S. Knight[2], Gustau Catalan[3], Simon A.T. Redfern[3], James F. Scott[3,4], Philip Lightfoot[1] and Finlay D. Morrison[1*]

[1] School of Chemistry, University of St Andrews, North Haugh, St Andrews, Fife, KY16 9ST, UK
[2] ISIS Facility, Rutherford Appleton Laboratory, Chilton, Didcot, OX11 0QX, UK
[3] Department of Earth Sciences, University of Cambridge, Downing Street, Cambridge, CB2 3EQ, UK.
[4] Department of Physics, University of Cambridge, J. J. Thomson Ave., Cambridge, CB3 0HE, UK
* To whom correspondence should be addressed: e-mail: fm40@st-andrews.ac.uk, Tel: +44-1334463855, Fax: +44-1334463805.

## Supplementary Information

**Table S1:** Parameters derived from Rietveld refinement of PND data as a function of temperature. Both β-BiFeO$_3$ and γ-BiFeO$_3$ were modelled in the orthorhombic space group Pbnm. Only lattice parameters were refined due to data quality with the exception of the 945 °C data as described in table 1 in the main text and marked (†) below.

| T (°C) | $\chi^2$ | wRp (%) | Rp (%) | a (Å) | b (Å) | c (Å) | Cell Vol (Å$^3$) | % BiFeO$_3$ |
|---|---|---|---|---|---|---|---|---|
| 910 | 0.976 | 7.81 | 7.87 | 5.6322(3) | 5.6547(1) | 7.9882(2) | 254.41(2) | 95 |
| 910 | 0.902 | 7.53 | 7.53 | 5.6319(3) | 5.6542(1) | 7.9875(2) | 254.36(1) | 93 |
| 920 | 0.901 | 7.50 | 7.46 | 5.6338(1) | 5.6547(2) | 7.9889(2) | 254.51(1) | 85 |
| 920 | 0.945 | 7.65 | 7.60 | 5.6328(2) | 5.6543(2) | 7.9886(2) | 254.43(1) | 78 |
| 930 | 0.942 | 7.69 | 7.65 | 5.6337(2) | 5.6550(2) | 7.9894(2) | 254.53(1) | 66 |
| 930 | 0.947 | 7.82 | 7.77 | 5.6317(2) | 5.6539(2) | 7.9880(2) | 254.35(1) | 56 |
| 935 | 0.923 | 7.74 | 7.78 | 5.6313(2) | 5.6538(2) | 7.9876(2) | 254.31(1) | 46 |
| 935 | 0.993 | 8.07 | 7.93 | 5.6306(2) | 5.6533(2) | 7.9866(2) | 254.23(1) | 41 |
| 940 | 0.971 | 8.09 | 7.89 | 5.6310(2) | 5.6535(2) | 7.9871(2) | 254.27(1) | 34 |
| 940 | 0.942 | 7.80 | 7.71 | 5.6306(2) | 5.6532(2) | 7.9870(2) | 254.23(1) | 33 |
| 945 | 1.153 | 8.49 | 8.06 | 5.6309(2) | 5.6535(2) | 7.9863(3) | 254.24(2) | 30 |
| 945† | 0.989 | 8.20 | 7.93 | 5.6311(3) | 5.6527(2) | 7.9878(3) | 254.26(2) | 27 |
| 950 | 0.918 | 7.88 | 7.69 | 5.6315(2) | 5.6536(3) | 7.9884(3) | 254.34(2) | 26 |
| 950 | 0.912 | 8.00 | 7.87 | 5.6317(3) | 5.6533(3) | 7.9879(3) | 254.31(2) | 22 |

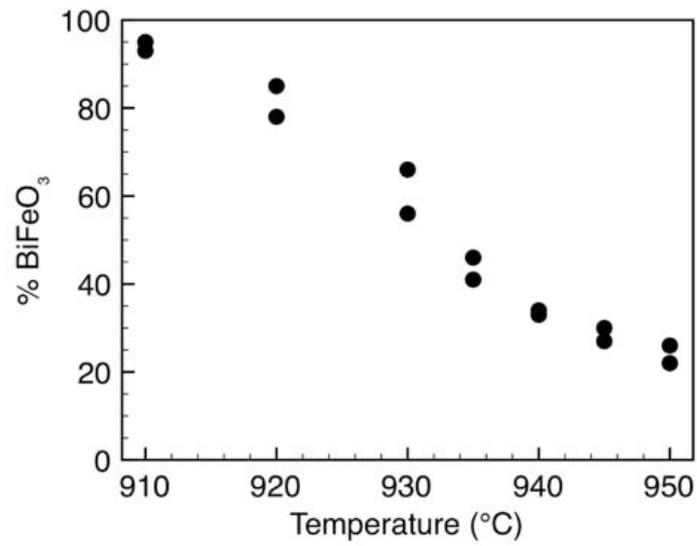

**Figure S1.** Percentage perovskite phase BiFeO$_3$ present as a function of temperature as determined from multi-phase Rietveld refinement of PND data. Two data points at each temperature represent first and second runs, as shown in table S1.

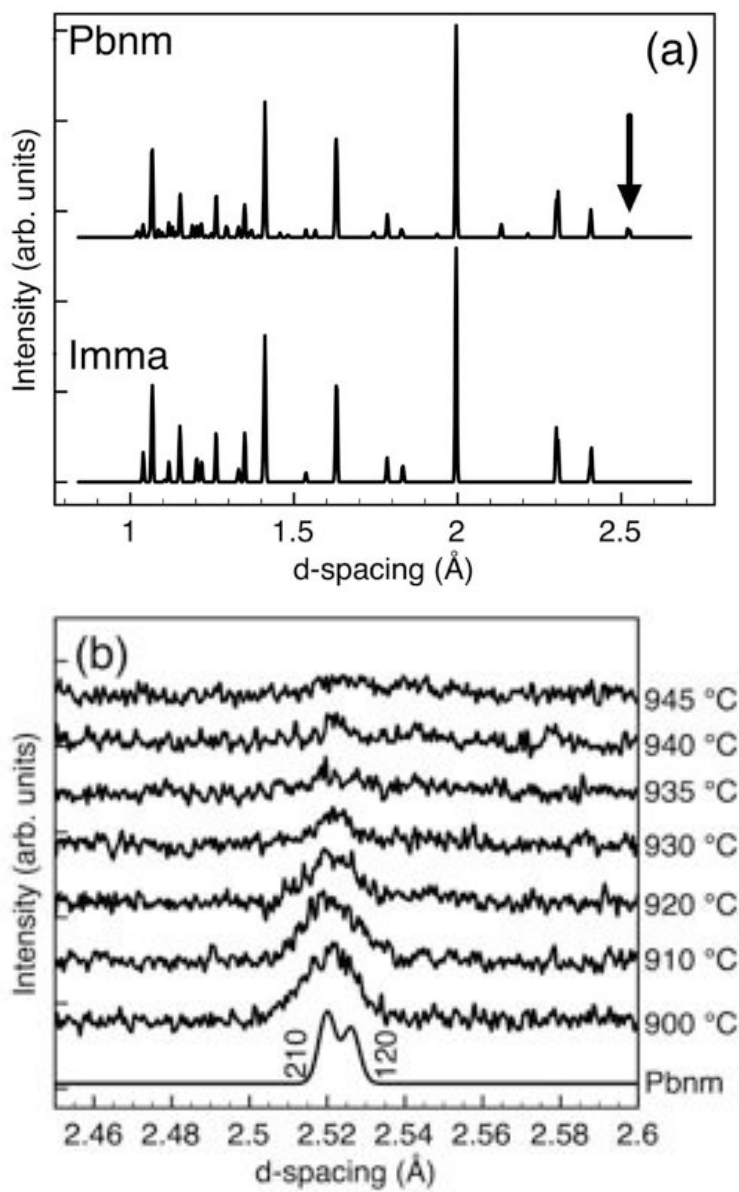

**Figure S2.** (a) Simulated PND patterns at 945 °C for Pbnm and Imma orthorhombic symmetry indicating additional reflections in Pbnm, including the (210/120) reflection (marked by arrow); (b) PND data in the region 2.45-2.60 Å as a function of temperature showing the presence of reflections consistent with Pbnm symmetry.

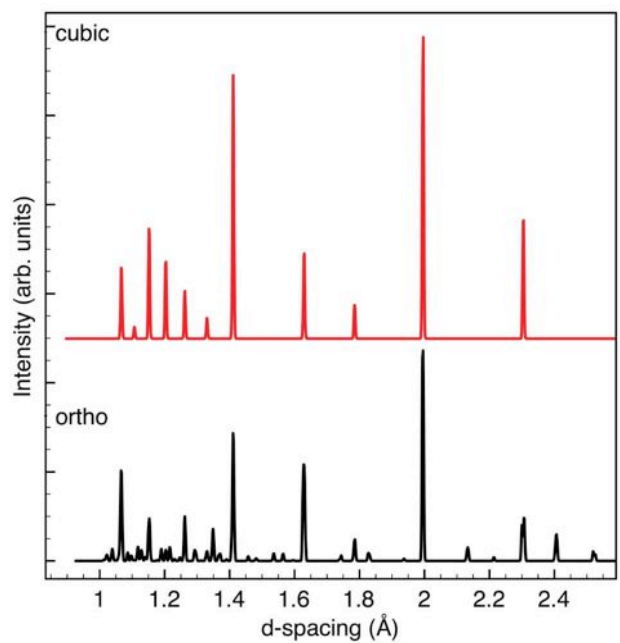

**Figure S3.** Simulated PND pattern showing the differences between cubic (Pm-3m, top) and orthorhombic (Pbnm, bottom) symmetries emphasising the clear ability of PND to differentiate between models, in comparison with figure 5a in the main text.

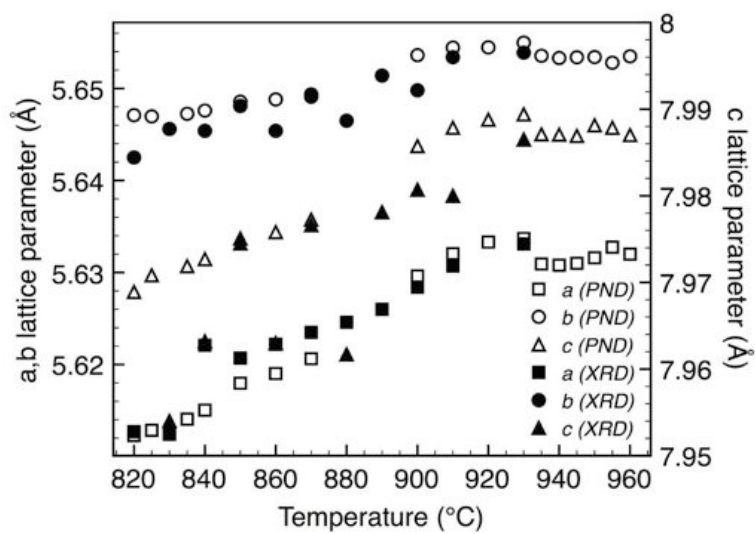

**Figure S4.** Comparison of lattice parameters obtained from data fitting of x-ray (XRD) and neutron (PND) diffraction data to the Pbnm space group.